\documentclass[twocolumn,amsmath,amssymb,superscriptaddress,footinbib]{revtex4-2}

\usepackage{graphicx}
\usepackage{hyperref}
\usepackage{braket}
\usepackage{layouts}
\usepackage{siunitx}
\usepackage{epstopdf} 
\usepackage[english]{babel}
\usepackage[svgnames]{xcolor}
\usepackage{ulem}
\newcommand{\extdatafig}[1]{Extended Data Figure~\ref{#1}}

\begin{document}
\rmfamily
\raggedbottom

\title{Observation of quadrature-dependent unidirectional flow of light}

\newcommand{\Stanford}{E. L. Ginzton Laboratory, Stanford University, Stanford, CA 94305, USA.}
\newcommand{\Technion}{Department of Electrical and Computer Engineering, Technion – Israel Institute of Technology, 32000 Haifa, Israel.}
\newcommand{\MIT}{Present address: LIGO, Massachusetts Institute of Technology, Cambridge, Massachusetts, USA}
\newcommand{\Harvard}{Present address: Brightlight Photonics, Cambridge, MA, USA}

\author{Jamison Sloan}
\thanks{These authors contributed equally to this work.}
\affiliation{\Stanford}

\author{Eran Lustig}
\thanks{These authors contributed equally to this work.}
\affiliation{\Technion}

\author{Louise Schul}
\thanks{These authors contributed equally to this work.}
\affiliation{\Stanford}

\author{Trung Kien Le}
\affiliation{\Stanford}

\author{Melissa A. Guidry}
\affiliation{\Stanford}
\affiliation{\MIT}

\author{Daniil M. Lukin}
\affiliation{\Stanford}
\affiliation{\Harvard}

\author{Shanhui Fan}
\affiliation{\Stanford}

\author{Jelena Vu\v{c}kovi\'{c}}
\affiliation{\Stanford}

\date[]{\textit{E-mail: jsloan@stanford.edu, elustig@technion.ac.il}}

\begin{abstract}
    Unidirectional transport of light is a central capability in photonics, enabling critical functionalities such as isolation and routing. 
    Ordinarily, this directionality depends on the mode of light, but not its phase. 
    In this work, we experimentally demonstrate phase-dependent unidirectional transport of light for the first time. Specifically, we realize a system where conjugate quadratures of an optical field decouple and propagate in opposite directions. 
    Our all-optical approach uses multiple pump frequencies in a high-$Q$ nanophotonic micro-resonator to achieve Hamiltonian programming through Kerr nonlinearity. 
    Even though we rely on optical nonlinearity, our approach is not limited to any minimal power requirement on the signal. We demonstrate the ultimate limit of this by measuring the effect of quadrature-dependent transport on vacuum fluctuations. 
    Our approach unlocks a fundamental new capability for controlling light, and can enable previously inaccessible regimes of phase-sensitive directional amplification, sensing, and routing.
\end{abstract}

\maketitle 

\section*{Introduction}

\begin{figure*}[htbp]
    \centering
    \includegraphics{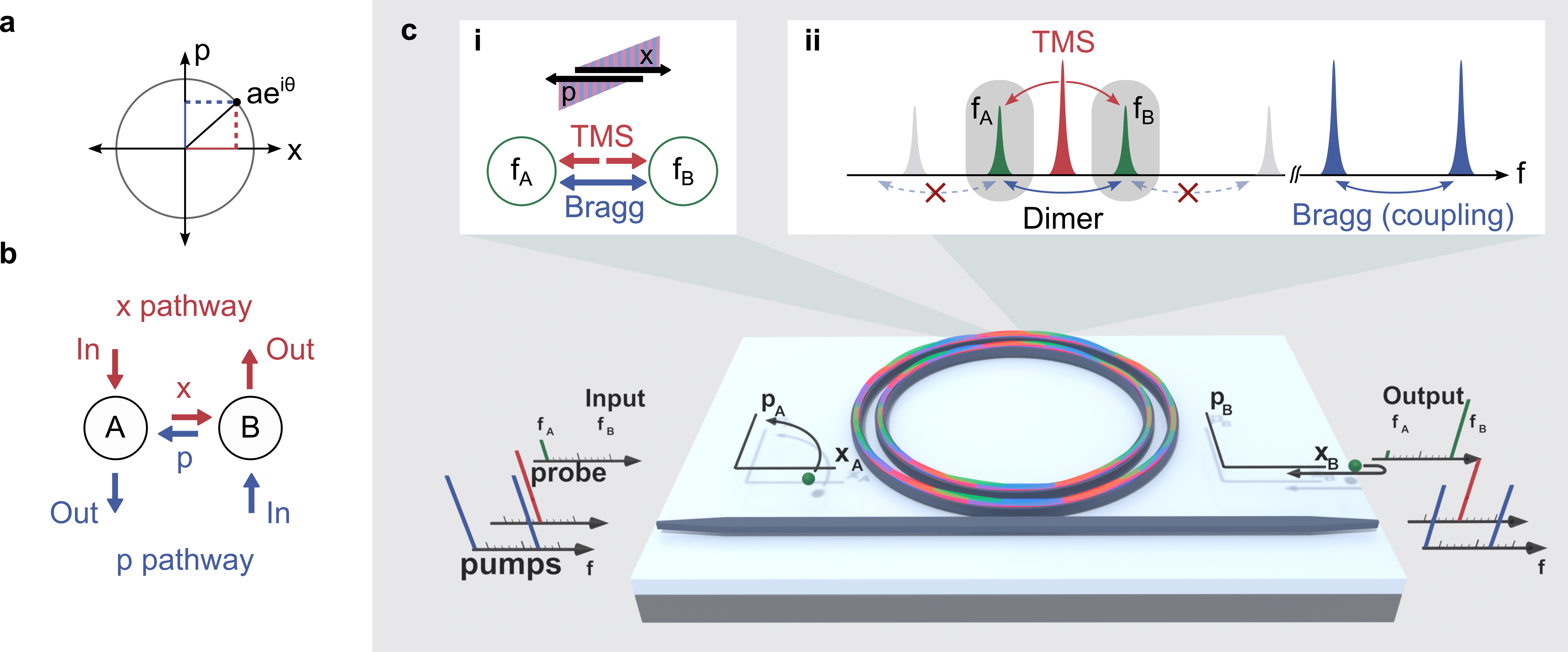}
    \caption{
    \textbf{Concept and implementation of quadrature-dependent directional transport between two optical modes.} 
    (\textbf{a}) A mode of light with a complex amplitude can be decomposed into two conjugate quadratures $x$ and $p$ as $|a|e^{i\theta} = x+ip$. 
    (\textbf{b}) Quadrature-dependent directionality between two modes of light, $A$ and $B$, occurs when light in the $x$ quadrature flows preferentially from $A \to B$, while light in the $p$ quadrature flows preferentially from $B \to A$. In the ideal case of \textit{unidirectionality}, the pathways are completely decoupled.
    (\textbf{c}) A multi-frequency pump field drives a high-$Q$ micro-resonator, producing simultaneous two-mode squeezing (TMS, magenta) and Bragg scattering (blue) between a selected pair of resonator modes at frequencies $f_A$ and $f_B$. The interference between these phase-coherent nonlinear processes results in phase-sensitive directional amplification, realizing quadrature-dependent directionality (inset i). In the ideal case of unidirectionality, an input probe at $f_A$ with variable phase will be amplified into $f_B$ only in quadrature $x_B$. Inset ii shows how the three tones generate the necessary interactions. Dispersion engineering of the device makes Bragg scattering to undesired modes non-resonant, restricting the interactions to the two mode ``dimer.''   
    }
    \label{fig:concept}
\end{figure*}

The unidirectional flow of light is fundamental to many areas of classical and quantum optics such as high power amplification \cite{white_unified_2024}, routing \cite{wang_observation_2009,rechtsman_photonic_2013,peng_paritytime-symmetric_2014} and lasing \cite{bahari_nonreciprocal_2017,bandres_topological_2018}. Traditional unidirectional transport involves breaking Lorentz reciprocity, inhibiting all light from scattering backwards \cite{jalas_what_2013}. However, in many cases, and particularly in photonic integrated circuits, non-reciprocity is not available as an intrinsic property of the guiding material \cite{srinivasan_review_2022}, and must instead be engineered through nonlinearity \cite{white_integrated_2023}, temporal modulation \cite{yu2009complete, herrmann_mirror_2022}, or non-reciprocal dissipation \cite{PhysRevX.5.021025,PhysRevX.7.031001}. Each of these approaches carries significant operational trade-offs, such as the requirement for high signal powers or fast electronic currents.

These challenges have motivated the pursuit of unidirectional transport without breaking Lorentz reciprocity. Such approaches exploit the fact that a reciprocal system can separate into two decoupled sub-systems, each supporting unidirectional transport; reciprocity relates each sub-system to its partner rather than constraining either alone. The unidirectionality is thereby locked to a degree of freedom, such as polarization \cite{petersen_chiral_2014,ma_all-si_2016}, orbital angular momentum \cite{wu_scheme_2015}, or propagation direction \cite{hafezi_imaging_2013}, and persists as long as the decoupling is maintained. While these approaches cannot achieve full optical isolation, they can be useful, and in some cases \cite{zhao_non-hermitian_2019,bandres_topological_2018} outperform systems that break Lorentz reciprocity. They can also operate in linear, time-reversal symmetric systems, and support counter-propagating unidirectional channels. 

All of these approaches to achieve unidirectionality for light rely on distinguishable modal degrees of freedom. A single optical mode, however, possesses a phase space described by the two conjugate quadratures $x$ and $p$ (Fig.~\ref{fig:concept}a). This permits a particularly interesting form of unidirectionality, which is locked to the phase of light \cite{mcdonald_phase-dependent_2018,wanjura_quadrature_2023}: light propagates one way for $x$ and the opposite way for $p$. This \textit{quadrature-dependent unidirectionality} requires interference between linear scattering (beamsplitting) and parametric gain (two-mode squeezing) processes across multiple modes. Figure~\ref{fig:concept}b shows the simplest manifestation in a system with two modes $(A, B)$ coupled to external input-output ports. Here, the $x$ quadrature is routed and amplified from $A_\text{in}$ to $B_\text{out}$, while the $p$ quadrature is routed and amplified in the opposite direction, from $B_\text{in}$ to $A_\text{out}$. Meanwhile, a $p$-quadrature signal entering $A_\text{in}$ has zero transmission to $B_\text{out}$, and an $x$-quadrature signal entering $B_{\text{in}}$ has zero transmission to $A_{\text{out}}$.

Quadrature-dependent unidirectionality is a fundamental building block for several important capabilities. Even with two modes, it enables intracavity squeezing beyond the 3 dB limit of traditional parametric oscillators \cite{dassonneville_dissipative_2021,11190781,8qtt-symt}. Moreover, ``bosonic Kitaev chains" \cite{mcdonald_phase-dependent_2018,wanjura_topological_2020} are lattices that exhibit this behavior and can enable exponentially enhanced sensing \cite{mcdonald_exponentially-enhanced_2020}. These prospects have motivated recent remarkable demonstrations in superconducting circuits \cite{busnaina_quantum_2024} and in the nanomechanical modes of optomechanical networks \cite{wanjura_quadrature_2023, slim_optomechanical_2024}. Despite these advances, quadrature-dependent unidirectionality has still never been demonstrated at optical frequencies. Such a capability would unlock a new class of phase-sensitive routing, sensing, and squeezing schemes for light.

Here, we experimentally demonstrate quadrature-dependent unidirectional flow of optical light. We observe this effect in a photonic integrated high-$Q$ micro-resonator utilizing an all-optical approach which leverages Kerr nonlinearity. We demonstrate complete control and dynamic tunability over the unidirectional transport, achieved by modifying the optical pump spectrum in a dispersion-engineered micro-molecule that prevents parasitic scattering - a major challenge \cite{zhang_squeezed_2021} in implementing unidirectional quadrature-dependent transport which we overcome in this work.
Additionally, we observe that the quadrature-dependent directionality leads to phase sensitivity and mode asymmetry in the two-mode squeezing produced by the device. Even though our approach is based on nonlinear optical interactions, the signal itself has no minimal power requirements for unidirectional behavior, and is therefore compatible with manipulation of low power states for sensing and quantum applications. Our approach is also fully integrated and CMOS compatible, making it readily available for usage in various applications.

\section*{Scheme for quadrature-dependent unidirectionality in optics}

\begin{figure*}[ht]
    \centering
    \includegraphics{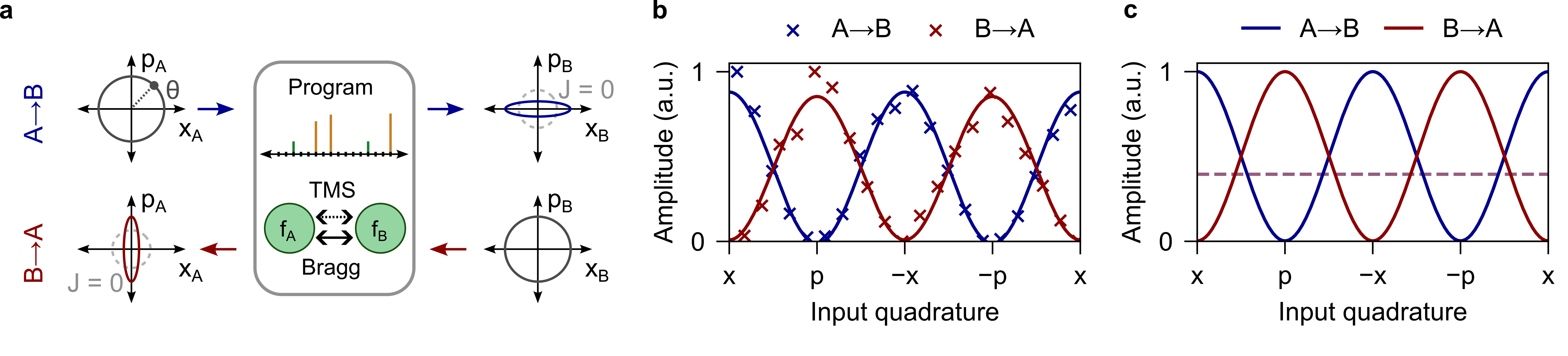}
    \caption{
    \textbf{Demonstration of quadrature-dependent unidirectionality of light.}
    (\textbf{a}) Measurement protocol for extracting the classical quadrature response. A coherent probe is prepared at quadrature angle $\theta$, the device is programmed with a selected Bragg and/or two-mode-squeezing (TMS) interaction, and the output quadrature is measured. Scanning $\theta$ maps how the programmed interaction rotates, amplifies, or de-amplifies the probe in phase space.
    (\textbf{b}) Experimentally measured quadrature-dependent unidirectionality. The blue $A\to B$ trace corresponds to injection into mode $\mu=+5$ and detection at $\mu=-5$, while the red $B\to A$ trace corresponds to the reverse process. The phase origins are chosen such that $A\to B$ transmission is maximal for $x_A$ and $B\to A$ transmission is maximal for $p_B$. Solid lines are best fits to the interference pattern stated in Eq.~\ref{eq:interference}.
    (\textbf{c}) Theoretically expected transmission behavior $T(\theta)$ for perfect quadrature-dependent unidirectionality, using the same quadrature convention as in panel b. It shows perfect extinction for the conjugate input quadratures in both directions. The dashed lines show that no quadrature dependence occurs without Bragg scattering ($J=0$).
}
    \label{fig:classical_probe}
\end{figure*}

We first briefly introduce the notion of quadrature-dependent flow of light. In a passive linear system, the scattering behavior of light has no global preferred phase: changing the phase of an input channel will simply shift the phase of all output channels. However when nonlinear parametric processes are present, the dynamics of light can become sensitive to phase, and thus exhibit preferred quadratures. Moreover, the interference of passive scattering and parametric processes can generate quadrature-dependent directional flow for light (Fig. \ref{fig:concept}b). A simple model which realizes this type of behavior consists of two modes of light, which are coupled through a coherent combination of two different optical processes: two-mode-squeezing (TMS), and beamsplitter-like scattering between the modes (Fig.~\ref{fig:concept}c, inset i). In this case, the dynamics of light in the two modes are described by the Hamiltonian:
\begin{equation}
    H/\hbar = \Delta_A a^\dagger a + \Delta_B b^\dagger b + \left(\lambda ab  + J a^\dagger b + \text{h.c.}\right).
    \label{eq:H}
\end{equation}
Here, $a$ and $b$ are the annihilation operators for the two modes, $\Delta_A$ ($\Delta_B$) is the detuning of mode $A$ ($B$), $\lambda$ produces two-mode-squeezing, and $J$ produces linear scattering. We refer to this model as the ``Dimer'' model.

This model is known to exhibit quadrature-dependent (i.e., phase-sensitive) directional transport \cite{mcdonald_phase-dependent_2018, wanjura_quadrature_2023}. In general, quadrature transport in a pair of modes is described by a quadrature scattering matrix $S$ which acts on quadratures $\mathbf{R}=(x_A, p_A, x_B, p_B)^T$ as $\mathbf{R}_\text{out} = S \mathbf{R}_\text{in}$. The scattering matrix $S$ is derived from input-output theory in a dissipative system (see S.I.). To explain the unidirectional behavior, we consider the case when detunings $\Delta_A=\Delta_B=0$ are zero, $\lambda$ and $J$ are both purely imaginary. Then, when light enters mode $A$ in the $x$ quadrature, it is preferentially transferred to $B$, and light entering mode $B$ in the $p$-quadrature is preferentially transferred to $A$; meanwhile, the conjugate pathways are suppressed. When TMS and linear transport interfere perfectly ($|J|=|\lambda|$), the system exhibits ``complete" unidirectional transport in which the $x$ and $p$ channels completely decouple. In this case, $p$-quadrature light entering $A$ has zero transmission to $B$, and $x$-quadrature light entering mode $B$ has zero transmission to $A$. 
This notion can be defined more generally for our system by demanding the nullification of the determinant of the scattering matrix, which is a gauge invariant property (see supplementary note).

We now describe our all-optical approach and experimental setup for generating and controlling quadrature-dependent directional transport for optical light. Specifically, we realize the Hamiltonian of Eq. \ref{eq:H} for two optical modes, and utilize tunability of $\lambda$ and $J$ to reach complete unidirectional transport. In our approach, the modes $A$ and $B$ are resonance frequencies $f_A$ and $f_B$ of a nanophotonic micro-molecule resonator (Fig.~\ref{fig:concept}c). Our micro-molecule consists of two concentric high-Q ($Q\approx1.5$ million) single mode rings of 4H-SiC on SiO$_2$ insulator \cite{lukin_4h-silicon-carbide--insulator_2020}. Both interactions (linear scattering and TMS) are realized by individually controlling the spectrum lines of the multi-frequency electro-optic comb pump (repetition rate $\approx 20$ GHz) which induces these interactions through the strong Kerr nonlinearity of the device (see inset ii). 

First, the TMS is driven by pumping a resonance at ~1541 nm (red pump in Fig. \ref{fig:concept}c) which sits between modes $A$ (1539.4 nm) and $B$ (1542.6 nm). We measured $0.45\pm0.06~dB$ of two-mode squeezing below the shot-noise level outside the chip, together with $2.77\pm0.04 ~dB$ of antisqueezing (see Methods and supplementary for the noise statistics and the detection loss budget) \cite{dutt_-chip_2015}, and observed optical parametric oscillation (OPO) threshold crossing in these modes for injected power of $\approx$30 mW. Second, the linear scattering between modes $A$ and $B$ is driven by simultaneously pumping an additional and distinct pair of cavity resonances (1543.5 nm and 1540.4 nm) with frequency difference $\Delta f = f_B - f_A$. These can in principle be any pair of resonances with appropriate frequency spacing. When these frequency-tones are present, Kerr nonlinearity drives Bragg scattering between modes $A$ and $B$ \cite{li_efficient_2016}. This approach enables us to achieve tunable scattering between different frequency modes all-optically, without applying electro-optic modulation \cite{herrmann_mirror_2022}. To tune the ratio between TMS and Bragg scattering, we simply change the spectrum of the pump. Additional theoretical details about how the effective two-mode Hamiltonian emerges from the full dynamics of the Kerr ring are provided in the S.I.
 
The last ingredient of our scheme is the controlled isolation of linear scattering and TMS for two optical modes ($A$ and $B$) which is enabled by the device geometry. Specifically, a concentric double-ring (micro-molecule) exhibits a sharp mode crossing that provides strong local anomalous dispersion with each individual ring having normal dispersion \cite{zhang_squeezed_2021}. This anomalous dispersion allows for efficient TMS over a narrow spectral window. The sharp mode crossing also introduces a large frequency mismatch with modes at adjacent frequencies, strongly suppressing Bragg scattering into these neighboring modes (depicted in Fig.~\ref{fig:concept}c, inset ii and \extdatafig{fig:extended-dispersion}b). This approach allows for the isolated control of scattering between two modes in a Kerr micro-resonator (which under normal circumstances generates interactions between many modes) \cite{guidry_quantum_2022,lustig_quadrature-dependent_2025}.

\section*{Demonstration of quadrature-dependent unidirectionality}

The characteristic signature of quadrature-dependent directionality is that the total power delivered from one port to the other varies with the input phase of an injected probe, rather than remaining flat. To demonstrate this behavior, we inject a weak probe with controllable phase $\theta$ into one mode ($A$ or $B$), and then measure the power transmission $T(\theta)$ to the opposite port (Fig. \ref{fig:classical_probe}a). For scattering from $A \to B$, the output field has the general form $b_{\text{out}} = U_{ba} a_{\text{in}} + V_{ba} a_{\text{in}}^*$. Here, $U_{ba}$ arises purely from Bragg scattering, and $V_{ba}$ arises purely from TMS. 
Then, a probe $a_{\text{in}} = |a_{\text{in}}|e^{i\theta}$ produces the $\theta$-dependent power transmission
\begin{equation}
    T(\theta) = |U_{ba}|^2 + |V_{ba}|^2 + 2 |U_{ba}V_{ba}^*| \cos(2\theta - \varphi),
    \label{eq:interference}
\end{equation}
where $\varphi=\arg(U_{ba}V_{ba}^*)$ is a possible phase offset and $T$ is normalized to the input power. This makes it clear that both processes must be present simultaneously ($|U_{ba}V_{ba}| \neq 0)$ in order to produce phase-dependent transmission. In the schematic (Fig. \ref{fig:classical_probe}a), the input field with variable phase traces out a circle, which is then converted into an ellipse in quadrature space in the output mode. Complete contrast is achievable when the two processes interfere with matched magnitudes ($|\lambda| = |J|$).

By programming our device to match the strength of TMS and Bragg ($|J| = |\lambda|$), we experimentally achieve the conditions required for quadrature-dependent unidirectionality. We observe that both the $A \to B$ and $B \to A$ responses exhibit high-contrast sinusoidal oscillations as the input quadrature is varied via $\theta$, in close agreement with theory (Figs.~\ref{fig:classical_probe}b,c). Indeed, we see that injecting $x_A$ or $p_B$ gives maximal power transmission to the other port, while injecting $x_B$ or $p_A$ results in almost no transmission. Nearly complete contrast is reached (i.e., for two phases, the detected power approaches zero). This observation confirms that Bragg and TMS interfere coherently and with comparable strengths, generating the observed quadrature-dependent transport. This vanishing of cross-port transport in both directions, for particular injected quadratures, is in fact equivalent to the condition that $|J| = |\lambda|$, even in the presence of detunings and asymmetric modal losses (see S.I. for additional discussion). Thus, our system is not only phase sensitive, but realizes quadrature-dependent unidirectionality.

\begin{figure}[!t]
    \centering
    \includegraphics{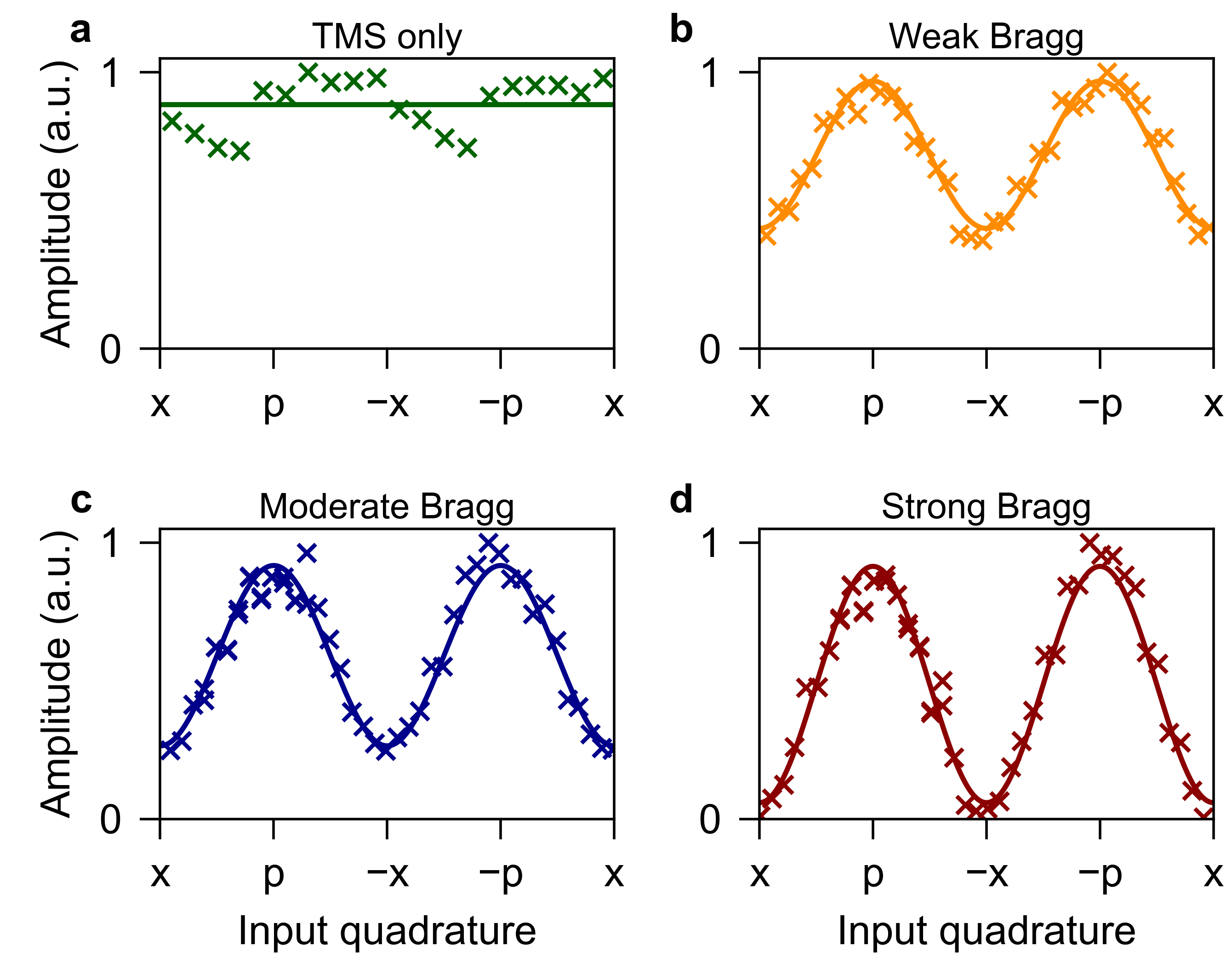}
    \caption{
    \textbf{Tunability of quadrature-dependent directionality.}
    Phase-dependent transmission from mode $B \to A$ for different levels of Bragg scattering strength $J$. The input phase is referenced to the convention of Fig.~\ref{fig:classical_probe}, such that the phase-sensitive transmission maxima in panels (\textbf{b-d}) occur for the $p_B$ input quadrature.
    (\textbf{a}) With TMS only, the response remains near the reference level set by the line.
    (\textbf{b-d}) Introduction of Bragg coupling produces a sinusoidal quadrature dependence, with a contrast that increases with the strength of $J$. At the highest level of Bragg strength (\textbf{d}), Bragg and TMS levels are matched, resulting in nearly perfect extinction in the phase-sensitive amplification.
    }
    \label{fig:classical_probe_vary_bragg}
\end{figure}

The all-optical programmability of our platform further enables us to tune the system through different levels of quadrature directionality (Fig. \ref{fig:classical_probe_vary_bragg}). In particular, by varying the strength of the Bragg term relative to the TMS, we demonstrate varying levels of phase-sensitive interference in the device. As the strength of Bragg scattering relative to TMS increases, the contrast in the phase-dependent power oscillations increases, consistent with theory. Moreover, for the strongest Bragg scattering, nearly complete contrast is again reached, corresponding to unidirectional quadrature transport.  

\section*{Quadrature directionality of amplified quantum noise}

\begin{figure*}[ht]
    \centering
    \includegraphics{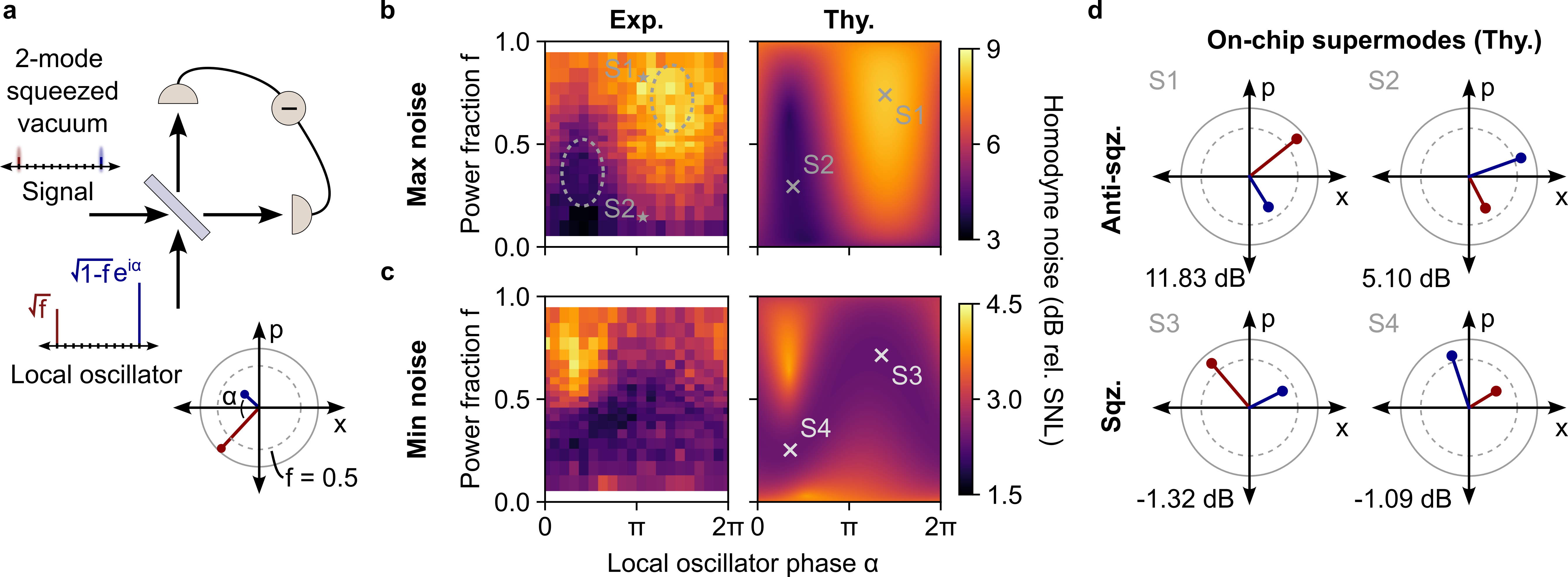}
    \caption{
    \textbf{Observation of quadrature-dependent directional amplification of vacuum fluctuations.}
    (\textbf{a}) Schematic of the two-mode homodyne measurement performed on amplified vacuum state from the programmed dimer. The signal is interfered with a two-component local oscillator (LO), whose relative phase $\alpha$ and power fraction $f$ are controlled to project onto different collective quadratures of the two optical modes.
    (\textbf{b}) Measured antisqueezing (Exp.) as a function of LO phase $\alpha$ and LO power fraction $f$ compared with the best-fit two-mode dimer model (Thy.) evaluated at a 12~MHz analysis frequency. The color scale gives the antisqueezed quadrature noise in dB relative to shot noise. In the experiment, we highlight the two regions corresponding to the two antisqueezed supermodes $S1_\star$ and $S2_\star$. We highlight the corresponding supermodes in the model as $S1$ and $S2$. 
    (\textbf{c}) Measured minimum noise (Exp.) and theoretical model (Thy.) over the same local oscillator scan corresponding to (b). 
    (\textbf{d}) Antisqueezed ($S1$, $S2$) and squeezed ($S3$, $S4$) supermodes on-chip (i.e., without detection loss and technical noise) of the theoretical model corresponding to the experiment. Red and blue vectors show the complex weights of the two optical modes in each supermode; labels give the corresponding quadrature noise eigenvalues in dB relative to shot noise. The two dominant antisqueezed supermodes are strongly asymmetric in LO power fraction, consistent with the structure observed in the full LO scan. The predicted squeezing values are smaller than the antisqueezing values due to a combination of undercoupling and detection loss.
    Model parameters: coupling losses
    $\gamma_{\mathrm{c}}=(0.162,0.115)\times10^{9}\,\mathrm{rad\,s^{-1}}$,
    intrinsic losses
    $\gamma_{\mathrm{i}}=(0.423,0.343)\times10^{9}\,\mathrm{rad\,s^{-1}}$,
    detunings
    $\Delta=(0.029,0.538)\times10^{9}\,\mathrm{rad\,s^{-1}}$,
    Bragg coupling
    $J=0.254\times10^{9}\,\mathrm{rad\,s^{-1}}$,
    nonlinear coupling
    $\lambda=0.574\times10^{9}\,\mathrm{rad\,s^{-1}}$,
    Bragg reference phase $\phi=3.30$ rad,
    LO phase offset $\alpha_{\mathrm{offset}}=0.220$ rad,
    detection efficiency $\eta_{\mathrm{det}}=0.340$,
    and uniform excess noise $N=0.811$ shot-noise variance units.
    }
    \label{fig:quantum_noise_supermodes}
\end{figure*}

A notable feature of our approach is that it provides quadrature-dependent directionality even for very weak light. In this section, we demonstrate the ultimate limit of this capability, by measuring the effect of directional quadrature transport on vacuum fluctuations. To do this, we program the dimer model, but turn off the coherent probes. As a result, the quadrature-dependent transformation we established in the previous section acts on the input vacuum state of the dimer modes, and generates a two-mode Gaussian state which contains the noise signature of the directionally amplified quadrature dynamics.

We use balanced homodyne detection (Fig.~\ref{fig:quantum_noise_supermodes}a) to characterize the two-mode gaussian state generated by the dimer, which is described by a quadrature covariance matrix $\sigma_{ij} = \frac{1}{2}\braket{R_iR_j + R_jR_i}$, where $R = (x_A, p_A, x_B, p_B)^T$. Interference of the dimer output with a two-mode local oscillator (LO) projects $\sigma$ along a quadrature vector $z$ set by the LO, giving $(\Delta X_{\mathrm{LO}})^2 = z^T \sigma z$. The LO is fully tunable, and parameterized compactly in complex coordinates as $b_{\mathrm{LO}} = e^{i\theta}(\sqrt{f},\, \sqrt{1-f}\,e^{i\alpha})^T$, where $f$ is the power fraction in mode $A$, $\alpha$ is the relative phase between the two LO components, and $\theta$ is the global LO phase. 
In the experiment, we scan $f$ and $\alpha$ in steps while continuously scanning $\theta$, and record the maximum and minimum measured quadrature noise (Fig.~\ref{fig:quantum_noise_supermodes}b-c). 

\begin{figure*}[ht]
    \centering
    \includegraphics{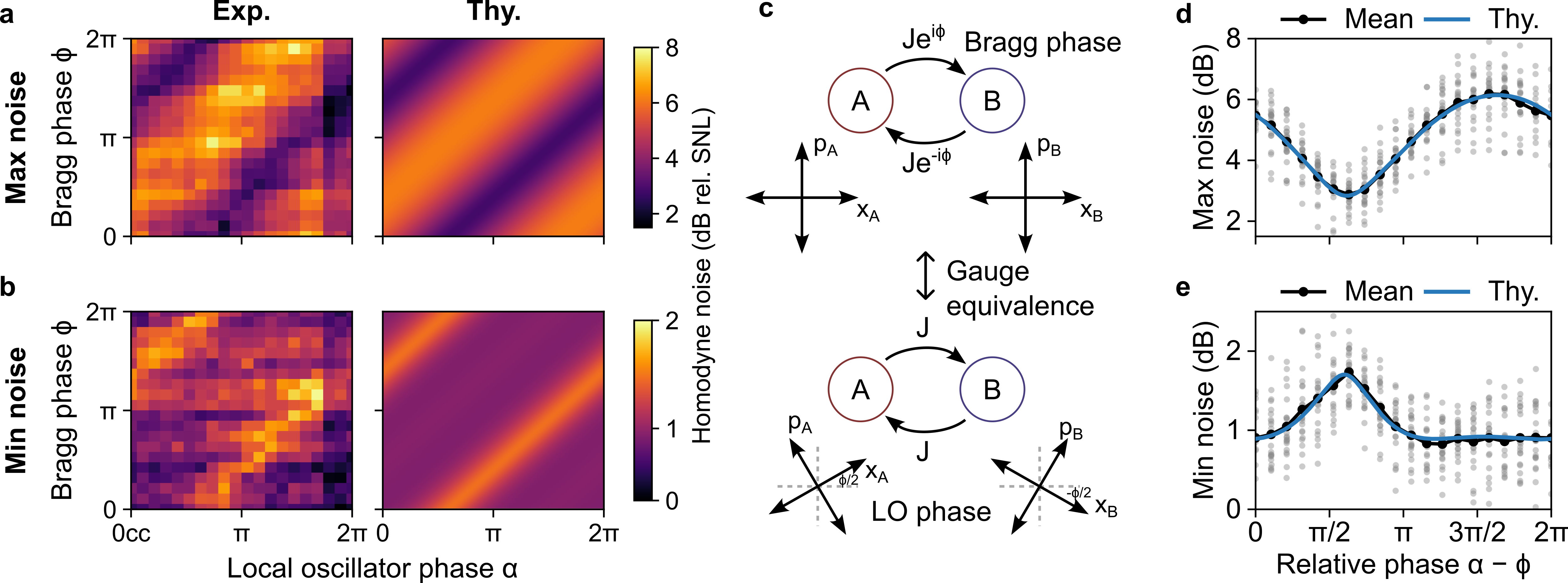}
    \caption{
    \textbf{Control over the basis of quadrature directionality for amplified vacuum fluctuations.}
    (\textbf{a}) Maximum homodyne variance measured (Exp.) and predicted (Thy.) with varying local oscillator phase $\alpha$ and Bragg phase $\phi$, at a fixed LO power fraction $f=0.7$ (where maximal antisqueezing is observed). Homodyne measurement is done at RF frequency of 12 MHz. 
    (\textbf{c-d}) Same as (a-b), for the minimum homodyne noise with varying local oscillator and Bragg phases. 
    (\textbf{c}) Schematic illustration of the phase gauge freedom in the two-mode dimer. A complex Bragg coupling $|J| e^{i\phi}$ can be transformed to a real coupling $J$ by rotating the quadrature axes of the two modes in opposite directions. In this real gauge, the phase previously carried by the Bragg coupling appears as an equivalent relative rotation of the local-oscillator quadratures.
    (\textbf{d-e}) Collapse of the phase-scan data onto the effective relative phase $\alpha-\phi$ after correcting for column-dependent vertical drifts (which we attribute to waveshaper leakage). Gray points show the drift-corrected measurements, the black curve shows the experimental mean, and the blue curve shows the model evaluated at the same power fraction. The agreement confirms that the measured phase dependence is governed mainly by the gauge-invariant relative phase between the LO and the dimer coupling.
    Parameters used for the theoretical model are: detunings
    $\Delta=(-0.788,0.011)\times10^{9}\,\mathrm{rad\,s^{-1}}$,
    Bragg coupling $J=0.238\times10^{9}\,\mathrm{rad\,s^{-1}}$,
    nonlinear coupling $\lambda=0.649\times10^{9}\,\mathrm{rad\,s^{-1}}$,
    Bragg phase $\phi=3.30\,\mathrm{rad}$, LO phase offset
    $\alpha_{\mathrm{offset}}=-4.012\,\mathrm{rad}$, detection efficiency
    $\eta_{\mathrm{det}}=0.340$, and uniform excess background noise
    $N=0.309$ shot-noise variance units.
    }
    \label{fig:quantum_noise_guage_covariance}
\end{figure*}

The first important feature of the measured fluctuations (Fig.~\ref{fig:quantum_noise_supermodes}b-c) is that they depend strongly on the relative mode phase $\alpha$. This is a direct result of the interference between TMS and Bragg scattering that produced quadrature-dependent directionality. Such behavior fundamentally differs from that of the two-mode squeezed state generated by conventional non-degenerate parametric oscillation (in the absence of the additional Bragg scattering), which cannot show any $\alpha$ dependence. Physically, the quadrature directionality produced by the interference of TMS and Bragg scattering generates amplification in a preferred quadrature direction, corresponding to homodyne noise levels that depend on the measured basis. This phase sensitivity is well matched by our theoretical model of squeezing dynamics in the dimer (see S.I. for additional details).

The second important feature of the measured fluctuations is that they directly reveal the preferred basis of the directional quadrature dynamics. In particular, they provide a direct observation of the collective ``supermodes'' of the amplified vacuum state, which correspond to eigenvectors of the covariance matrix $\sigma$. A general two mode squeezed state has two antisqueezed supermodes, with noise above the vacuum level, and two squeezed supermodes, with noise below the vacuum level. In our experiment, the most strongly anti-squeezed supermode ($S1_\star$) is observed around $f_1 \approx 0.7$ (Fig.~\ref{fig:quantum_noise_supermodes}b). Importantly, this maximum is shifted away from equal mode weight ($f=0.5$), indicating that the dominant antisqueezed fluctuation is not shared equally between the two optical modes. The second antisqueezed supermode ($S2_\star$) appears as a stationary point in the antisqueezing measurements, and is observed at a power fraction which complements the first supermode (i.e., $f_2 \approx 0.3 \approx 1 - f_1$). Similarly, we observe variations in the minimum quadrature variance measured with varying local oscillator (Fig.~\ref{fig:quantum_noise_supermodes}c).

Our theoretical model reproduces the LO dependent experimental noise measurements well. The models for both the maximum and minimum measured quadrature variance includes both technical noise and the amplified vacuum. 
Further analytical theory allows us to explain these observations in more detail. Specifically, our theory shows that the presence of detunings $\Delta_i$ in our system breaks supermode degeneracy, resulting in two supermodes whose power fractions are complementary, $f_2 \approx 1-f_1$ (Fig.~\ref{fig:quantum_noise_supermodes}c). 
 Additional analysis of the quantum noise dynamics in the dimer, including the role of asymmetric coupling and intrinsic losses, is provided in the S.I.

As a final demonstration of the programmable control of quadrature directionality in our device, we show that the preferred quadratures which are amplified can be continuously tuned by varying the phase of the Bragg coupling $J = |J|e^{i\phi}$. This is achieved in the experiment simply by changing the phase of one of the auxiliary pump tones which generates Bragg scattering. 
We then independently vary both $\alpha$ and $\phi$ at a fixed LO power balance of $f=0.7$, and record the maximum and minimum noise to track the supermode behavior. As the Bragg phase is varied, the relative LO phase $\alpha$ which achieves the maximal antisqueezing peak shifts, as the natural basis of quadrature directionality is varied (Fig.~\ref{fig:quantum_noise_guage_covariance}a). Similar behavior is observed in the minimum quadrature variance (Fig.~\ref{fig:quantum_noise_guage_covariance}c).

This behavior is well reproduced by the same type of theoretical model used in Fig.~\ref{fig:quantum_noise_supermodes}. To understand this effect, we note that the Bragg phase $\phi$ can be removed from the dimer Hamiltonian (Eq.~\ref{eq:H}) by a phase rotation of the dimer modes: $a \rightarrow a e^{i\phi/2},\, b \rightarrow b e^{-i\phi/2}$. Shifting the Bragg phase $\phi$ is therefore equivalent to rotating the relative quadrature basis in which the two-mode covariance matrix is expressed (Fig.~\ref{fig:quantum_noise_guage_covariance}c). This interpretation is further confirmed by examining the measured noise as a function of the phase difference $\alpha-\phi$ (Fig.~\ref{fig:quantum_noise_guage_covariance}d-e). After correcting for slow column-dependent drifts in the noise level, the data collapse onto a single curve when plotted against $\alpha - \phi$. This shows directly that the Bragg phase controls the basis of directional quadrature flow. We note that this change of quadrature basis is mathematically equivalent to a gauge transformation on the dimer Hamiltonian (Eq.~\ref{eq:H}). Thus, our data can be equivalently interpreted as experimental verification that the quantum noise response depends only on the gauge-invariant relative phase.

\section*{Conclusion}

In summary, we have demonstrated quadrature-dependent unidirectional transport between two optical modes in an integrated Kerr nonlinear device. 
A central feature of this approach is that it is based on Hamiltonian programming of a single physical device. The Kerr nonlinearity provides a flexible mechanism for synthesizing both photon-conserving and photon-nonconserving processes using optical pumps, and the same strategy can in principle be extended from a two-mode dimer to larger multimode lattices \cite{lustig_quadrature-dependent_2025}. This opens a route toward optical implementations of more complex quadrature-directional bosonic systems, including bosonic Kitaev chains and related non-Hermitian sensing architectures, purely based on optical nonlinearity. Furthermore, integrated optics is a particularly natural setting for these ideas, since nanofabrication allows for dispersion engineering of individual modes, and the high field confinement enables low power operation.

More broadly, our work establishes quadrature-dependent routing as a new building block for controlling classical and quantum optical signals. Because the interaction is linear for the weak signal fields, the same device response applies to arbitrarily weak inputs, including single photons and other weak nonclassical states of light. This makes the platform relevant not only for phase-sensitive classical signal processing, but also for multimode quantum state generation, frequency conversion, and phase-sensitive routing. Such unidirectional quadrature interactions also underlie back-action-evading measurement schemes which could now be implemented in integrated photonics \cite{doi:10.1126/science.abf5389}. Another promising direction is to use quadrature-directional dynamics to engineer intracavity squeezing beyond the conventional 3 dB limit of a degenerate optical parametric oscillator. If implemented in cavities that also interface with quantum emitters, such as color centers or other solid-state defects, this approach could provide a path toward driving cavity-QED systems with strongly squeezed intracavity light. All of these directions suggest that programmable Kerr interactions in integrated photonics can provide a valuable toolkit for controlling the transport, amplification, and quantum fluctuations of light.

\section*{Methods}

\subsection*{Device}

Our device is a micro-molecule realized in a 4H-silicon carbide (4H-SiC) integrated platform~\cite{lukin_4h-silicon-carbide--insulator_2020}, consisting of two evanescently coupled concentric microring resonators.
The coupled-ring geometry supports two families of supermodes - symmetric (lower) and antisymmetric (upper) superpositions of the individual ring modes - whose resonance frequencies repel at a wavelength-dependent mode crossing.
This mode crossing is the key structural feature enabling our experiment, as it simultaneously provides the anomalous dispersion required for efficient parametric pair generation and the anharmonic mode spacing that localizes the Bragg scattering interaction to a single pair of modes.

Both rings are etched from a 300-nm-thick 4H-SiC-on-insulator wafer and are 2670 nm wide (inner) and 2025 nm wide (outer), with a ring diameter of 920 $\mu$m (outer) and 915 $\mu$m (inner), yielding a single-ring free spectral range of $\mathrm{FSR} \approx 39.2$~GHz.
The inter-ring coupling gap is 450 nm. The outer ring is point coupled to a waveguide (gap of 300 nm).
The rings are capped with SiO$_2$ for protection and mode confinement.
A resistive microheater, patterned via lift-off of platinum with a  titanium adhesion layer, is integrated on the chip directly above the outer ring and allows continuous thermal tuning of the resonances by up to two FSR, enabling the pump mode to be tuned from the cold resonance at $\lambda_0=1540.28$~nm to the operating wavelength $\approx 1541$~nm. We also utilize the micro-heater for Pound-Drever-Hall locking to stabilize the phase.

\extdatafig{fig:extended-dispersion} shows the measured integrated dispersion
\begin{equation}
    D_\mathrm{int}(\mu) = \omega_\mu - \left(\omega_0 + \mu D_1\right),
\end{equation}
where $\omega_\mu$ is the resonance frequency of the $\mu$-th mode, $\omega_0$ is the pump mode frequency, and $D_1/2\pi = \mathrm{FSR} \approx 39.2$~GHz is the free spectral range.
The dispersion was measured on the cold device; during the experiment, a microheater integrated on the chip shifts the resonances by one FSR so that the pump mode is thermally tuned to $\lambda_0 \approx 1540.6$~nm.

The OPO signal and idler modes are located at $\mu = +5$ ($\lambda_{+5} \approx 1538.7$~nm, cold) and $\mu = -5$ ($\lambda_{-5} \approx 1541.8$~nm, cold), both directly measured on the upper supermode branch with $D_\mathrm{int}(+5)/2\pi \approx +66$~MHz and $D_\mathrm{int}(-5)/2\pi \approx +178$~MHz.
The net phase mismatch $[D_\mathrm{int}(+5) + D_\mathrm{int}(-5)]/2\pi \approx +244$~MHz is compensated by the nonlinear cross-phase-modulation shift from the pump field, enabling above-threshold OPO at accessible pump powers of approx 30mW input power.
A polynomial fit to the upper branch for $|\mu|\leq 11$ yields $D_2/2\pi \approx 9.3$~MHz, $D_3/2\pi \approx -2.1$~MHz, $D_4/2\pi \approx -0.20$~MHz, and $D_5/2\pi \approx 0.27$~MHz.

The mode crossing also strongly suppresses parasitic scattering to modes $\mu = \pm 10$.
A linear ruler defined by the two OPO modes ($\mu = \pm5$) and extended by one additional spacing of 5 modes predicts $D_\mathrm{int}(-10)/2\pi \approx 234$~MHz and $D_\mathrm{int}(+10)/2\pi \approx 9$~MHz.
The measured values are $496$~MHz and $259$~MHz respectively, deviating from the ruler by $+262$~MHz and $+250$~MHz (\extdatafig{fig:extended-dispersion}b).
These large deviations (compared to the linewidths) place modes $\pm10$ far off-resonance with the Bragg pump pair tuned for the $\pm5$ dimer, effectively suppressing any parasitic cascade.

\extdatafig{fig:extended-dispersion}c shows the measured $Q$ factors of the upper supermode branch across all 89 measured modes.
At the pump mode ($\mu=0$) the loaded, coupling, and intrinsic quality factors are $Q_L \approx 1.0\times10^6$, $Q_c \approx 2.9\times10^6$, and $Q_i \approx 1.6\times10^6$, corresponding to linewidths $\kappa_L/2\pi \approx 188$~MHz, $\kappa_c/2\pi \approx 66$~MHz, and $\kappa_i/2\pi \approx 122$~MHz.
Since $\kappa_i > \kappa_c$, the device operates in the under-coupled regime, consistent with the measured off-chip squeezing of $0.45\pm0.06$~dB.
A prominent feature of the $Q$-factor spectrum is the rapid growth of $Q_c$ toward positive mode indices (shorter wavelengths, approaching the mode crossing): $Q_c$ increases from $\approx 2.9\times10^6$ at the pump to $\approx 8\times10^6$ at $\mu > 20$.
This reflects a progressive reduction in waveguide-to-resonator coupling as the upper supermode hybridizes with the lower branch near the crossing, making those modes increasingly dark.
By contrast, $Q_i$ remains relatively flat across the spectrum, indicating that the intrinsic loss of the resonator does not vary strongly with mode index.
The device therefore realizes an isolated photonic dimer: two modes coupled by independently controllable two-mode squeezing and Bragg scattering interactions, with no parasitic coupling to the surrounding multimode resonator spectrum.

\subsection*{Experimental setup}

The experiment is driven by a single continuous-wave laser source, which is used to generate both the pump fields and the optical local oscillator. The laser is first amplified using a pre-amplifier and an erbium-doped fiber amplifier (EDFA), and then sent through three cascaded stages of fiber electro-optic modulation to generate a frequency comb with 20 GHz line spacing. A programmable waveshaper separates this comb into pump and local-oscillator branches. This configuration allows the pump spectrum and the local-oscillator spectrum to be configured independently from the same phase-coherent set of comb lines.

In the pump branch, selected comb lines are amplified with an additional EDFA and sent to the device through lensed-fiber coupling. The output from the chip is collected with an objective lens and refocused into a fiber in order to maximize collection efficiency. Throughout the setup, we use a combination of tunable bandpass filters and fixed telecom channel filters to suppress EDFA amplified spontaneous emission and isolate the relevant pump, signal, and local-oscillator tones with high signal-to-noise ratio. The local-oscillator and signal branches are then interfered and sent to a balanced detector for either heterodyne or homodyne measurements, depending on the experiment.

Stable phase operation is essential because the measured response depends on optical quadrature. We therefore actively phase lock the pump and signal branches to maintain a stable phase relationship between the programmed Hamiltonian and the measurement basis. In addition, the device resonance frequency is stabilized during operation using an integrated microheater on the chip. This heater lock compensates slow thermal and environmental drifts of the resonator, allowing the nonlinear interaction and measurement detunings to remain stable over the duration of the experiment.

\subsection*{Single-pump OPO characterization}

Before programming the dimer with the full set of pumps, we characterize the bare parametric response of the device driven by a single continuous-wave pump and no Bragg tones.
In this configuration the molecule realizes a conventional optical parametric oscillator (OPO) that couples the two dimer modes $\mu=+5$ and $\mu=-5$ through four-wave mixing with the pump at $\mu=0$ ($\lambda_0\approx 1541$~nm).
This single-pump measurement isolates the two-mode squeezing interaction from the Bragg-scattering interaction, and serves two purposes: it quantifies the quantum noise performance of the device below threshold, and it verifies that the parametric gain is maximized at the measured dimer modes.

\emph{Below threshold.}
We first operate the pump below the OPO threshold and perform balanced homodyne detection on the amplified vacuum fluctuations of the $\mu=\pm5$ pair.
Scanning the local-oscillator (LO) phase rotates the measurement axis through the squeezing ellipse, so the measured noise power oscillates about the shot-noise level (\extdatafig{fig:extended-singlepump}a).
The anti-squeezed quadrature rises $2.77\pm0.04$~dB above shot
noise, while the orthogonal quadrature dips $0.45\pm0.06$~dB below
it. The measurement is taken at 30 MHz frequency, with resolution
bandwidth of 10 kHz, and video bandwidth of 10 Hz; these set a $0.19$~dB
statistical uncertainty per displayed point, matching the $0.18$~dB scatter measured on the shot-noise trace. Both levels are obtained by fitting the full local-oscillator scan rather than from trace extrema, and the quoted uncertainties are $1\sigma$. The recovered off-chip squeezing is modest because the device is under-coupled at the pump ($\kappa_i > \kappa_c$, see above) and is further reduced by the finite detection efficiency.

\emph{Above threshold.}
Increasing the pump power past threshold, the device oscillates and emits bright signal and idler tones.
\extdatafig{fig:extended-singlepump}b shows the output optical spectrum just above threshold: the only lines present are the pump ($\mu=0$) and the signal/idler pair at $\mu=\pm5$ ($\lambda\approx 1539.8$ and $1542.9$~nm, separated from the pump by $5\times\mathrm{FSR}\approx 196$~GHz), each $\sim\!17$~dB above the noise floor.
No other comb lines appear: the OPO energy stays entirely in the two dimer modes we measure.
At a higher pump power (\extdatafig{fig:extended-singlepump}c) the $\mu=\pm5$ lines grow, but the next potential comb order at $\mu=\pm10$ remains more than $30$~dB below the OPO lines and only marginally above the optical spectrum analyzer noise floor, and no cascaded broadband comb develops.
This strong suppression is a direct consequence of the same mode-crossing anharmonicity that isolates the Bragg interaction: modes $\mu=\pm10$ are pushed far off the frequency grid set by the $\mu=\pm5$ pair (\extdatafig{fig:extended-dispersion}b), so cascaded four-wave mixing into them is heavily phase-mismatched.
The device therefore behaves as a clean two-mode OPO rather than a frequency-comb generator, which is essential for interpreting the quadrature-transport measurements as a genuine two-mode process.
In Fig.~\ref{fig:classical_probe_vary_bragg}a, the TMS-only trace is not perfectly flat because it is not using extra band pass filtering on the pump as in the quantum squeezing measurements, showing a residual modulation of $\approx$20\% peak-to-peak at twice the input-phase frequency. This originates from the finite extinction of the waveshaper used to program
the pumps: the amount of light leaking into the driven ports depends on how the remaining, empty ports are programmed. Reprogramming only an unused port which cannot alter the state of the device changes the residual from 12.6\% contrast at 23$^\circ$ to 16.9\% at 171.5$^\circ$, whereas the Bragg-induced modulation retains a fixed phase (155--170$^\circ$) across the whole series. The residual therefore sets a floor on the contrast that can be resolved in this measurement. 

\subsection*{Classical probe measurements}

For the classical probe measurements, the waveshaper is used to prepare the optical tones that program the two-mode Hamiltonian, together with an additional weak tone that acts as the coherent probe. The pump tones generate the desired combination of two-mode squeezing and Bragg scattering between the selected dimer modes. The probe tone is injected into one of the dimer modes, and its phase is controlled directly with the waveshaper. By scanning this probe phase, we measure how different input quadratures are transferred and amplified by the programmed interaction.

The field generated in the opposite dimer mode is measured using a balanced heterodyne scheme. In this measurement, the detected mode frequency is prepared in the local-oscillator branch and shifted with an acousto-optic modulator (AOM). After interference between the local oscillator and the classical signal, the resulting AOM beat note is detected on a balanced receiver. Demodulating this beat note gives access to both the amplitude and phase of the transferred signal. This allows us to reconstruct the phase-sensitive classical response of the device as a function of the injected probe quadrature and programmed Bragg phase.

\subsection*{Two-mode homodyne measurements}

For the quantum-noise measurements, the waveshaper again prepares the pump tones that program the two-mode Hamiltonian, but no coherent signal is injected into either dimer mode. The two modes are therefore seeded only by vacuum fluctuations at the input. These vacuum fluctuations are amplified by the programmed Bragg--TMS interaction, producing a two-mode Gaussian output state. To measure the quadrature noise of this state, we generate an arbitrary two-mode local oscillator from the same electro-optic comb. The waveshaper controls both the relative power fraction and the relative phase of the two local-oscillator components, allowing us to project the output field onto different collective two-mode quadratures.

The two-mode local oscillator is interfered with the amplified vacuum field and detected on a balanced homodyne receiver. The resulting difference current is sent to an electrical spectrum analyzer, which measures the quadrature-noise power at the chosen radio-frequency analysis frequency. By scanning the global local-oscillator phase via a phase-stretcher, we extract the maximum quadrature noise for each choice of relative local-oscillator phase and power fraction. This procedure maps the antisqueezed covariance landscape of the two-mode output state.

\section*{Acknowledgments}\label{sec7}
We gratefully acknowledge discussions with Edwin Ng. This work is funded by the Defense Advanced Research Projects Agency under the QuICC program and by the NSF QuSeC-TAQS Award ID 2326792, as well as the Vannevar Bush Faculty Fellowship from the US Department of Defense and AFOSR Award number: FA9550-23-1-0248. E.L acknowledges the Yad Hanadiv Rothschild fellowship, and the Zuckerman institute Zuckerman fellowship. J.S. acknowledges the support of an Intelligence Community Postdoctoral fellowship. Part of this work was performed at the Stanford Nanofabrication Facility (SNF) and the Stanford Nano Shared Facilities (SNSF). S. F. acknowledge the support of a MURI project for the U. S. Air Force Office of Scientific Research (Grant No. FA9550-22-1-0339).  We thank NGK Insulators, Ltd. for 4H-SiCOI substrates used to fabricate devices in this work.

\bibliographystyle{apsrev4-2}
\bibliography{references.bib}

\newpage
\setcounter{figure}{0}
\renewcommand{\figurename}{Extended Data Figure}

\begin{figure*}
    \centering
    \includegraphics[width=\textwidth]{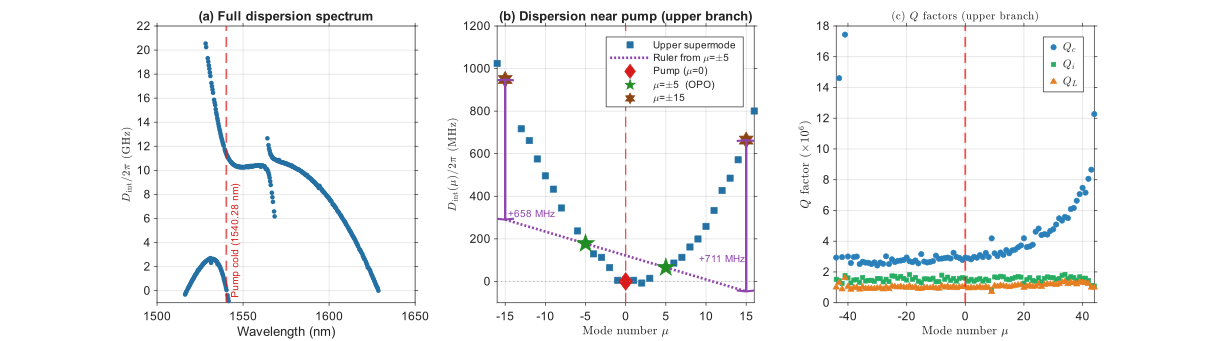}
    \caption{
    \textbf{Measured integrated dispersion of the concentric-ring micro-molecule.}
    (\textbf{a})~ Full dispersion spectrum $D_\mathrm{int}/2\pi$ versus wavelength for both supermode branches of the coupled-ring resonator. The lower supermode branch (branch below $D_\mathrm{int}/2\pi$ of 4GHz in y-axis and below 1550 nm in wavelength) exhibits a zero crossing near the pump wavelength $\lambda_0 \approx 1541$~nm (red dashed line) due to the hybridization with the upper branch at the mode crossing. A second zero crossing appears at
    longer wavelength; it originates from a separate intermodal crossing and does
    not involve the modes used in this work.
    (\textbf{b})~Integrated dispersion of the upper supermode branch relative to the pump ($\mu=0$, red diamond; cold $\lambda=1540.28$~nm, thermally tuned to $\approx1541$~nm). Both OPO modes ($\mu=\pm5$, green stars) are directly measured. The dotted purple line is a linear ruler defined by modes $\mu=\pm5$ and extended by one additional 5-mode spacing; the double-headed arrows show that modes $\mu=\pm15$ (brown stars) deviate from this ruler by $+262$ and $+250$~MHz respectively, placing them far off-resonance with the Bragg pump pair and suppressing parasitic scattering.
     (\textbf{c})~Measured $Q$ factors of the upper supermode branch across all 89 modes. The coupling $Q_c$ (blue) grows strongly toward negative $\mu$ as the modes approach the crossing and become increasingly dark, while the intrinsic $Q_i$ (green) remains relatively uniform. The red dashed line marks the pump mode ($\mu=0$).
    }
    \label{fig:extended-dispersion}
\end{figure*}
\newpage

\begin{figure*}
    \centering
    \includegraphics{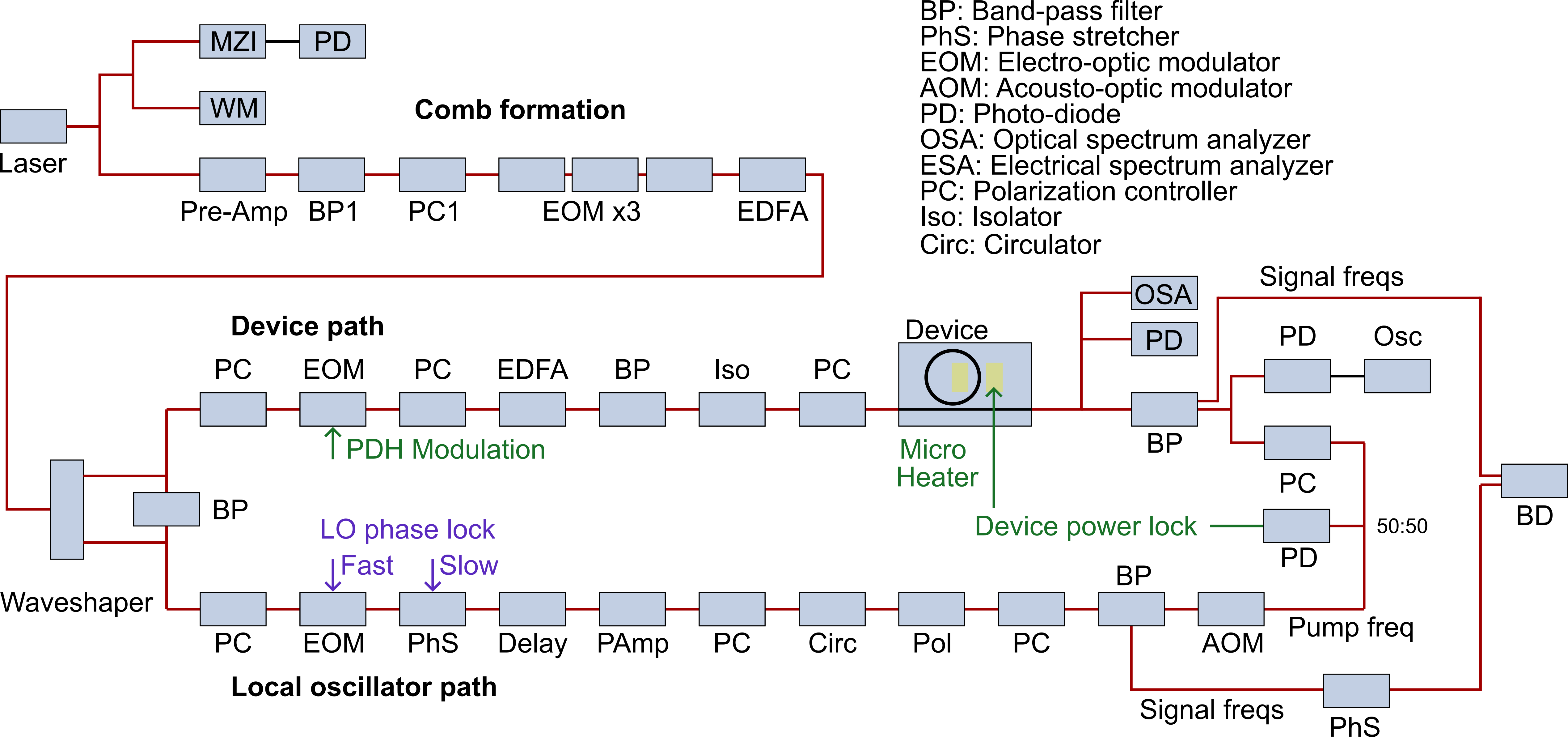}
    \caption{\textbf{Full experimental schematic.} A continuous wave laser is split into two paths: one is used for wavelength monitoring, and another is amplified before being sent into a series of 3 electro-optic (EO) modulators driven at 20 GHz to form a comb. The comb is sent into a waveshaper, which splits the comb into two paths: a device path, and a local oscillator (LO) path. The device path is amplified and sent into the integrated photonic device through a lensed fiber. Light collected from the device through an objective lens is interfered with the LO path, and sent into a balanced detector (BD). Both the device and LO paths contain modulators used for active locking control of the power and phase stability of operation.}
    \label{fig:extended-schematic}
\end{figure*}
\newpage

\begin{figure*}
    \centering
    \includegraphics[width=\textwidth]{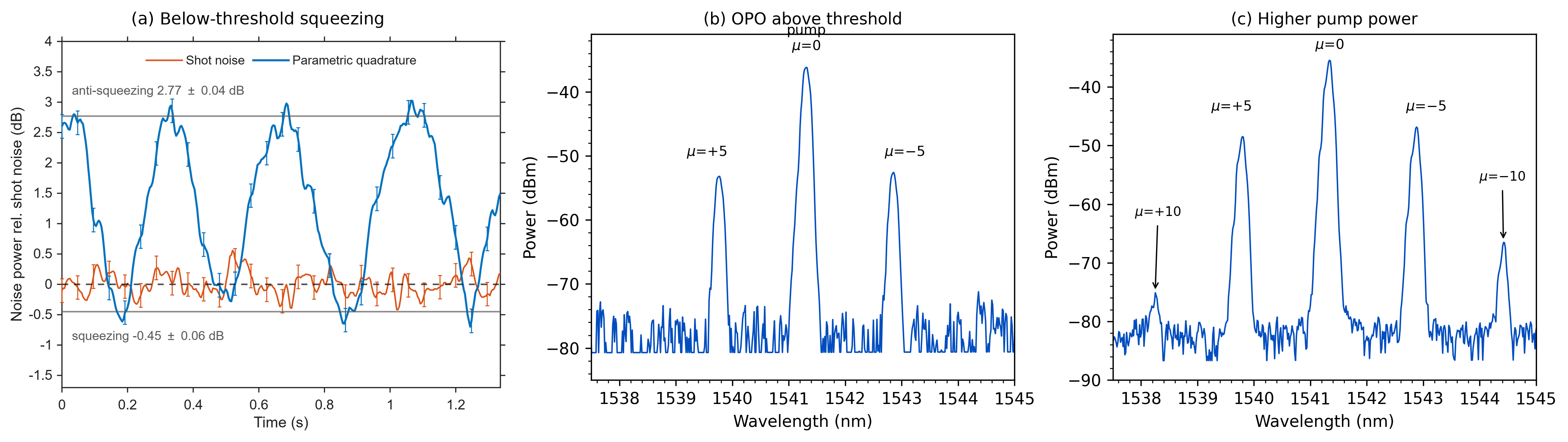}
    \caption{
    \textbf{Single-pump characterization of the OPO between the dimer modes.}
    (\textbf{a})~Below-threshold squeezing. Balanced-homodyne measurement of the $\mu=\pm5$ pair, seeded by vacuum, as the LO phase is scanned (blue), referenced to the shot-noise level (orange; dashed line at $0$~dB). The parametric quadrature is anti-squeezed by $2.77\pm0.04$~dB and its orthogonal quadrature dips $0.45\pm0.06$~dB below shot noise; error bars are the $\pm1\sigma$ statistical uncertainty per point.
    (\textbf{b})~Above-threshold output spectrum. Only the pump ($\mu=0$) and the signal/idler pair at $\mu=\pm5$ are present; the OPO oscillates exclusively in the two measured dimer modes with no comb formation.
    (\textbf{c})~Output spectrum at higher pump power. The $\mu=\pm5$ lines dominate while the next comb order $\mu=\pm10$ (arrows) stays $>30$~dB below them, confirming that the mode-crossing anharmonicity suppresses cascaded comb generation.
    }
    \label{fig:extended-singlepump}
\end{figure*}

\end{document}